# Intelligent smartphone-based portable network diagnostics for water security: Case Study – real-time pH mapping of tap water

Md. Arafat Hossain,[1,2] John Canning,[1,*] Sandra Ast,[1] Peter J. Rutledge,[1] and Abbas Jamalipour[2]

[1]*interdisciplinary Photonic Laboratory (iPL) & School of Chemistry, The University of Sydney, NSW 2006, Australia*
[2]*School of Electrical and Information Engineering, The University of Sydney, NSW 2006, Australia*
*\*Corresponding author: john.canning@sydney.edu.au*

Using a field-portable, smartphone fluorometer to assess water quality based on the pH response of a designer probe, a map of pH of public tap water sites has been obtained. A custom designed Android application digitally processed and mapped the results utilizing the GPS service of the smartphone. The map generated indicates no disruption in pH for all sites measured. All the data are assessed to fall inside the upper limit of local government regulations and are consistent with authority reported measurements. The work demonstrates a new security concept: environmental forensics utilizing the advantage of real-time analysis for the detection of potential water quality disruption at any point in the city. The concept can be extended on national and global scales to a wide variety of analytes.

OCIS Codes: (999.9999) Lab-in-a-phone, (280.4788) Optical sensing and sensor, (060.4261) Networks, protection and restoration, (999.9999) Smartphone sensing, (170.6280) Spectroscopy, fluorescence and luminescence;

Regular and real-time monitoring of drinking water quality is likely to become an essential feature of urban life given the potential for natural and deliberate contamination. Amongst various parameters used to measure water quality, pH is a particularly important assessment criterion. Aligned with the World Health Organization (WHO), the Australian government's National Health and Medical Research Council (NHMRC), for example, provides a generally acceptable alkaline range of pH ~ 6.5 to 8.5, which avoids skin corrosion at the acidic end and irritation of skin, eye and mucous membranes at the basic end [1]. The NHMRC mandate for pH monitoring at defined frequencies both at the supplier and consumer ends and average data is generally published on publicly accessible websites [2].

Other international agencies, such as the United States Environmental Protection Agency (EPA), are concerned with higher frequency monitoring because of the growing fear of hazardous disruption of water supplies by criminal, terrorist, military, industrial and environmental incursions [3-5]. Existing technologies monitor daily water quality at the supplier end [5], but less frequently at the consumer end since it is practically impossible at present to assess all points of access using current methods. It seems likely the growing local and global risk of water disruption will necessitate more frequent and a greater number of measurements than are made today. This process will be aided tremendously if the technology can be integrated into growing wireless sensor networks, which would allow rapid collation, detection and monitoring of propagation, site identification and response, ideally coordinated through an automated receiving data centre. Here, we propose smartphone-based sensor systems that can provide this capability. Such smartphone technology is increasingly used as a field-portable sensing platform [6-20]. Smartphones are becoming cheap with rapidly increasing market share [21] and there is the capacity for each instrument to communicate with the others as well as a central data processing node. This makes them ideal for use in networked systems. Mapping of heavy metal ion detection using two parts colorimetric analysis has been reported using a smartphone, [11] but this approach involved integration of two external diode sources and power supplies to image the samples, which greatly limits device accessibility for a large part of the world.

In this work, we advance our previous work on pH sensors [6] by introducing a reference cell in the attachment to allow both direct self-referencing and calibration. A customized Android application is reported that digitally processes the captured images and shares results along with the GPS coordinates of the corresponding locations. This allows effective real-time mapping of a region's water quality. The work reported here establishes the potential for novel forensic tools that can analyze this data to detect sudden disruptions in the network, such as the quality of water at any location. Using this "lab-in-a-phone" technology to measure pH, we have completed a case study of tap water across Sydney, the most common source of drinking water in this region. Many forms of disruption may push the pH of water outside of the recommended safe drinking levels [22] and the aim of this approach is to detect and demonstrate localization to a point of origin, using the mapping technique. The concept can be extended to measure a number of other parameters with an application-specific molecular probe on the same platform.

In this work, our portable fluorometer was deployed on an Android-driven Samsung Galaxy Express smartphone. The same fluorometer can also be implemented on other smartphones as well as other Android driven devices, such as Tablets, with slight mechanical modification. The

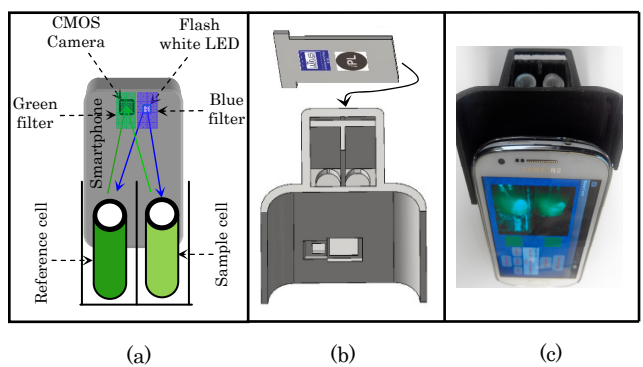

Fig. 1 (a) Schematic of the simple smartphone fluorometer design; (b) Smartphone attachment designed in AutoCAD Inventor Fusion, and (c) photograph from top view of the 3D printed attachment installed on an Android based smartphone.

schematic summarizing the principle of the smartphone diagnostic tool is shown in Fig. 1a. In this fluorometer, the source is the blue filtered white flash LED normally used to illuminate the background of an image whilst the camera itself is the imaging sensor. Use of this LED eliminates the need for any external sources as well as mirrors or other beam displacement hardware used in most smartphone based analytical systems [7]-[14], [20]. This in-built source, integrated with a suitable driver circuit internally connected to the battery, has substantially greater and more consistent irradiance and is also collocated with the camera, allowing optimal illumination during imaging. In order to further improve signal-to-noise ratios over our previous work [6], a green filter is also used during fluorescence imaging to remove any background scattering of blue light, a potential source of error. The 3D structure of the smartphone attachment that contains all the components (including the sample cells) was designed in Inventor software (AutoCAD) to fit over the top of the camera unit (Fig. 1b). This contains a sample and reference cell chamber and suitable slots for the color filters. The attachment was then fabricated in a low-cost MakerBot Replicator 2X 3D printer (Fig. 1c); the unit is robust for transport, keeps the sample well shielded from the outside, and excludes light from external sources.

The pH probe used in this work is the thermally stable, easily prepared, photo-induced electron transfer (PET) dye 4-aminonapthalimide (Fig. 2 inset). The absorption maximum of this probe aligns well with the wavelength of filtered light from the camera flash (absorption peak at ~444 nm with a 3dB bandwidth of ~70 nm fits with the source filtered emission peak at ~437 nm). The aminonaphthalimide dye functions as a turn-on probe in the presence of protons: the protonated molecule fluoresces strongly, but with rising pH, deprotonating of the dye leads to suppressed emission as a function of pH (up to ~pH 10, beyond which no further change in emission is observed). The relative fluorescence intensity of this probe across a range of pH with respect to a fixed pH solution is shown in Fig. 2. A customized, user-friendly Android application was developed to allow efficient pH measurements of tap water in the field, and sharing of re-sults (Fig. 3). After attaching the hardware to the camera unit of the smartphone, the user can hold the phone vertically and then run the measurements with this smart application. From the main menu of the application (Fig. 3a), the user can select to start a new test, create a device-specific calibration curve, view previously run tests, and review the operating instructions. After capturing the fluorescent image, the user can first preview the image on the screen (Fig. 3b) before proceeding to analyze it. The al-gorithm was designed to compare the fluorescent green intensity in a fixed region of 20 × 20 pixels in the sample image with respect to that from a fixed reference solution. Results from various sites can be stored on the phone's available memory (Fig. 3c) for mapping on the same platform (Fig. 3d) or transmitted back automatically with location names and GPS coordinates (Fig. 3e) to a central server for quick mapping, using, for example, a Google Maps-based interface [23].

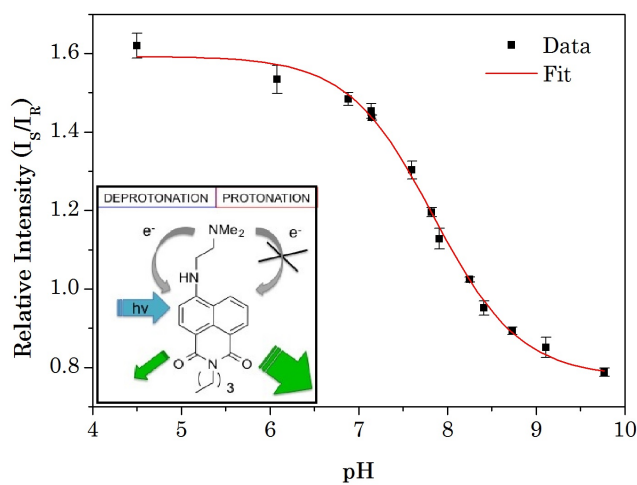

Fig. 2 Response curve of the pH probe, 2-butyl-6-((2(dimethyl-amino)ethyl)-mino)-1$H$-benzo[$de$] isoquino-line1,3(2$H$)-dione (inset: fluorescence switching mechanism) running on the smartphone system. The data are fitted by a logarithmic function with a co-efficient of determination ($R^2$) of 0.998.

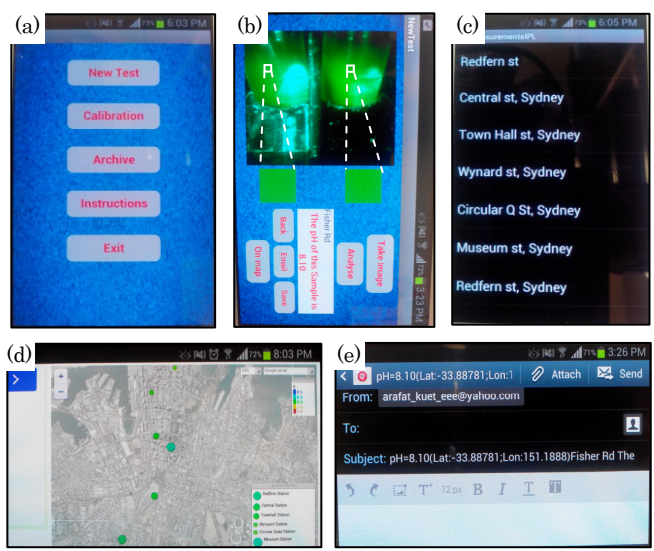

Fig. 3 Screenshots of the pH measurement application running on an Android phone: (a) The main menu; (b) The testing screen displaying results with location after analysis. Upon the selection, the results can be (c) stored on the phone's memory; (d) mapped with GPS coordinates on the same mobile platform (clear images are shown in Fig. 4); and (e) sending for quick, centralized mapping of results from many sources/instruments.

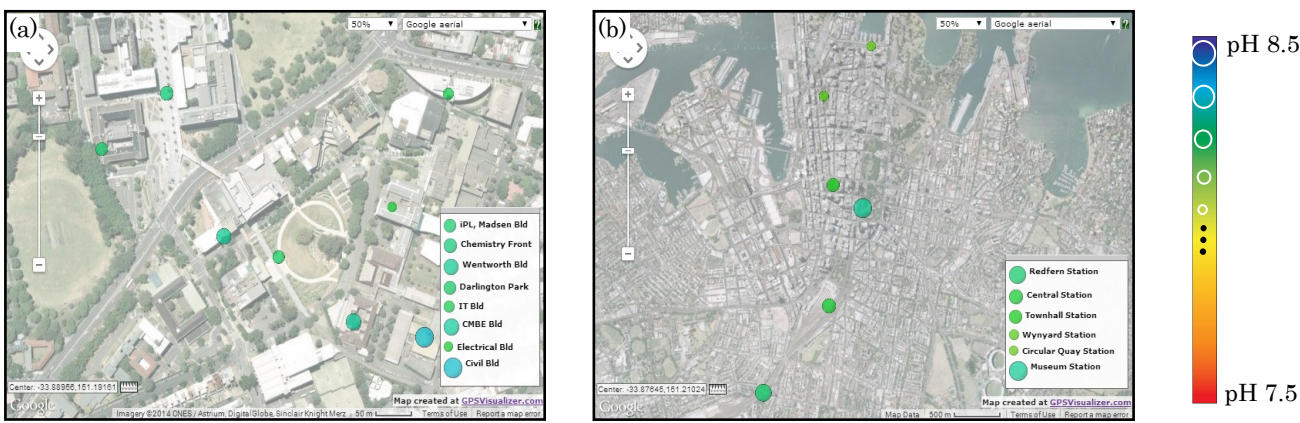

Fig 4. Results of tap water pH mapping at (a) The University of Sydney and (b) Sydney City Circle train stations. The color gradient from red to blue and dot size indicate the pH value.

In order to perform the measurements of unknown pH using our platform, the smartphone App was first calibrated against 13 standard samples ranging from pH 4.50 to 9.77 with the calibration option from the main menu (Fig. 3a). The samples were prepared as reported previously [6]. When all parameters (position of vials, dye concentration, volume, and excitation intensity) for both sample and the reference cell are identical, the relative fluorescence intensity can be correlated with the pH change in the water sample (Fig. 2). When the data are fitted to the Henderson–Hasselbalch equation, the calculated acid dissociation constant ($pK_a$ = 8.6) is very close to the reported values for this type of probe [24]; the small variation observed is presumably due to the different solvent system used. The calibration equation was then uploaded to the App to enable pH measurements of field samples.

Measurements of pH were conducted with samples from public tap water at different locations across Sydney, Australia. These included different buildings within The University of Sydney and different train stations around Sydney. In order to compare the results, the same samples were also measured with a commercial, portable pH meter (PHMETER, PH-035) which was calibrated with standard buffer solution before each measurement. Unlike the smartphone, this conventional system cannot be readily integrated into a wireless network and requires a distinct power supply that is not available at many sites globally. In contrast, the mobile fluorometer uses the same calibration procedure each time ensuring the instrument is truly field-portable whilst providing good agreement within small experimental error.

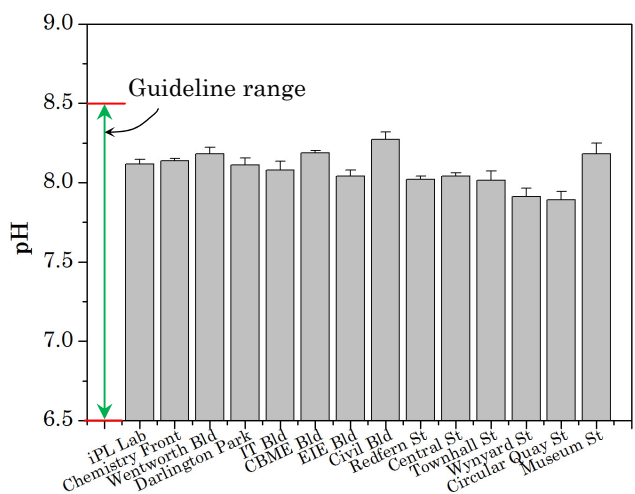

Fig 5. The histogram of measured pH at tap water sites in different buildings on The University of Sydney campus and Sydney City Circle train stations.

Notably, our smartphone based fluorometer can also generate real-time maps within the same platform, and share results with others for centralized mapping. To do this, GPS coordinates were recorded at each measurement location and saved within the phone's memory for mapping within the same device. The results were also sent wirelessly to a central computer at the *i*nterdisciplinary Photonic Laboratories (*i*PL) in The University of Sydney for quick mapping. This offers further advantages over the conventional pH meter, enabling direct integration into wireless sensor networks for direct access to central laboratory analysis. Fig. 4 represents two pH maps generated from smartphone fluorometer data sent from the recorded locations. The water within these locations is found to be slightly alkaline, with pH ~(7.89 - 8.27) ± 0.10 and an average $pH_{av}$ = 8.08 ± 0.10 (Fig. 5). The map generated for this region indicates no spikes in results, reflecting the fact that there was no disruption of pH in public drinking water at The University of Sydney or City Circle train stations during the period that these measurements were made. These results are also within the upper limit of the NSW government's tolerable alkalinity (pH = 8.5) and are consistent with the values most recently reported for this region ($pH_{av}$ = 8.10) [2]. In a fully implemented wireless sensor network, monitoring would be done on a regular basis and diagnostics and analysis automated at the data mapping centre. The monitoring range can be extended to a national and even a global scale by collecting similar data from many portable instruments (where resources permit, supplemented by permanent analytical systems), connected to the wireless infrastructure. This approach has great potential for monitoring many other environmental analytes, and for remote biological analysis on national and global scales.

In summary, we have proposed a new forensic concept - network and smartphone forensics - for environmental monitoring. This approach enables water security that can take advantage of next-generation wireless sensor technologies. We have demonstrated the application of a novel smartphone-based pH meter to rapidly assess and map the quality of drinking water around Sydney. The measured pH distribution is consistent at all sites with results published by official authorities using traditional technologies (and measured at fewer locations). As was expected for the reported regions, no evidence of water supply disruption was uncovered. Nonetheless, the approach can be scaled and introduced with relative ease. Combined with existing routine visits by water authority personnel and automated sensing using wireless technologies, this method can form the basis of an efficient network mapping tool for early detection of disruptions in water quality both in Australia and in more vulnerable regions of the world.

The authors acknowledge support from the Australian Research Council (ARC) through grants ARC FT110100116 and DP120104035. Md. Arafat Hossain acknowledges an International Postgraduate Research Scholarship (IPRS) from The University of Sydney.

### References

1. *Australian Drinking Water Guidelines 6, National Water Quality Management Strategy, Version 2,* page 174 (2013).
2. *Quarterly drinking water quality report, 1 Jul. 2013 to 30 Sep. 2013*, [online]: www.sydneywater.com.au/.
3. *The Potential Terrorist Risk of Drinking Water Contamination*, Office of Homeland Security & Preparedness Intelligence Bureau (Jan. 2009).
4. P. H. Gleick, Water Policy **8**, 481 (2006).
5. J. S. Hall, J. G. Szabo, S. Panguluri, and G. Meiners, U.S. Environmental Protection Agency, (Oct. 2009).
6. M. A. Hossain, J. Canning, T. L. Yen, S. Ast, P. J. Rutledge, and A. Jamalipour, OSA Congress in Optics and Photonics: Sensors, July 2014 [online]http://www.opticsinfobase.org/abstract.cfm?uri=Sensors-2014-SeTh2C.1.
7. D. N. Breslauer, R. N. Maamari, N. A. Switz, W. A. Lam, and D. A. Fletcher, PLoS ONE **4**, 1 (2009).
8. Z. J. Smith, K. Chu, A. R. Espenson, A. Gryshuk, M. Molinaro, D. M. Dwyre, S. Lane, D. Matthews, and S. Wachsmann-Hogiu, PLoS ONE **6**, 1 (2011).
9. H. Zhu, O. Yaglidere, T. Su, D. Tseng, and A. Ozcan, Lab Chip **11**, 315 (2011).
10. Q. Wei, H. Qi, W. Luo, D. Tseng, S. J. Ki, Z. Wan, Z. Gorocs, L. A. Bentolila, T. T. Wu, R. Sun, and A. Ozcan, ACS Nano **7**, 9147 (2013).
11. Q. Wei, R. Nagi, K. Sadeghi, S. Feng, E. Yan, S. J. Ki, R. Caire, D. Tseng, and A. Ozcan, ACS Nano **8,** 1121 (2014).
12. D. Gallegos, K. D. Long, H. Yu, P. P. Clark, Y. Lin, S. George, P. Natha, and B. T. Cunningham, Lab Chip **13**, 2124 (2013).
13. Y. Intaravannea, S. Sumriddetchkajorn, and J. Nukeawa, Sensors and Actuators B **168**, 390 (2012).
14. S. Sumriddetchkajorn, K. Chaitavon, and Y. Intaravanne, Sensors and Actuators B **191**, 561 (2014).
15. A. W. Martinez, S. T. Phillips, E. Carrilho, S. W. Thomas, H. Sindi, and G. M. Whitesides, Anal. Chem. **80**, 3699 (2008).
16. A. García, M. M. Erenas, E. D. Marinetto, C. A. Abada, I. O. Paya, A. J. Palma, and L. F. C. Vallvey, Sensors and Actuators B **156,** 350 (2011).
17. J. L. Delaney, E. H. Doeven, A. J. Harsant, and C. F. Hogan, Anal. Chim. Acta **790,** 56 (2013).
18. J. L. Delaney, C. F. Hogan, J. Tian, and W. Shen, Anal. Chem. **83**, 1300 (2011).
19. J. L. Delaney, E. H. Doeven, A. J. Harsant, and C. F. Hogan, Anal. Chem. Acta **790**, 56 (2013).
20. J. Canning, A. Lau, M. Naqshbandi, I. Petermann, and M. J. Crossley, Sensors **11**, 7055 (2011).
21. Mobile-cellular subscriptions 2013 [online]: http://www.itu.int/en/ITU-D/Statistics /Pages/stat/ default.aspx.
22. B. Oram, Water Research Centre, [online]: http://www.water-research.net/index.php/ph-in-the-environment.
23. *GPS Visualizer*, [online]: http://www.gpsvisualizer.com/.
24. A. P. de Silva, H. Q. N. Gunaratne, J. L. Habib-Jiwan, C. P. McCoy, T. E. Rice, and J.-P. Soumillion, Angew. Chem. Int. Ed. Eng. **34**, 1728 (1995).

# References (Full Version)

1. *Australian Drinking Water Guidelines 6, National Water Quality Management Strategy, Version 2,* page 174 (2013).
2. *Quarterly drinking water quality report, 1 Jul. 2013 to 30 Sep. 2013,* [online]: www.sydneywater.com.au/.
3. *The Potential Terrorist Risk of Drinking Water Contamination*, Office of Homeland Security & Preparedness Intelligence Bureau (Jan. 2009).
4. P. H. Gleick, "Water and terrorism" Water Policy **8**, 481 (2006).
5. J. S. Hall, J. G. Szabo, S. Panguluri, and G. Meiners, "Distribution system water quality monitoring: sensor technology evaluation methodology and results, a guide for sensor manufacturers and water utilities" U.S. Environmental Protection Agency, (Oct. 2009).
6. M. A. Hossain, J. Canning, T. L. Yen, S. Ast, P. J. Rutledge, and A. Jamalipour, "A smartphone fluorometer – the lab-in-a-phone" OSA Congress in Optics and Photonics: Sensors, July 2014 [online] http://www.opticsinfobase.org/abstract.cfm?uri=Sensors-2014-SeTh2C.1.
7. D. N. Breslauer, R. N. Maamari, N. A. Switz, W. A. Lam, and D. A. Fletcher, "Mobile phone based clinical microscopy for global health applications" PLoS ONE **4**, 1 (2009).
8. Z. J. Smith, K. Chu, A. R. Espenson, A. Gryshuk, M. Molinaro, D. M. Dwyre, S. Lane, D. Matthews, and S. Wachsmann-Hogiu, "Cell phone-based platform for biomedical device development and education applications" PLoS ONE **6**, 1 (2011).
9. H. Zhu, O. Yaglidere, T. Su, D. Tseng, and A. Ozcan, "Cost-effective and compact wide-field fluorescent imaging on a cell phone" Lab Chip **11**, 315 (2011).
10. Q. Wei, H. Qi, W. Luo, D. Tseng, S. J. Ki, Z. Wan, Z. Gorocs, L. A. Bentolila, T. T. Wu, R. Sun, and A. Ozcan, "Fluorescent imaging of single nanoparticles and viruses on a smart phone" ACS Nano **7**, 9147 (2013).
11. Q. Wei, R. Nagi, K. Sadeghi, S. Feng, E. Yan, S. J. Ki, R. Caire, D. Tseng, and A. Ozcan, "Detection and spatial mapping of mercury contamination in water samples using a smart-phone" ACS Nano **8**, 1121 (2014).
12. D. Gallegos, K. D. Long, H. Yu, P. P. Clark, Y. Lin, S. George, P. Natha, and B. T. Cunningham, "Label-free biodetection using a smartphone" Lab Chip **13**, 2124 (2013).
13. Y. Intaravannea, S. Sumriddetchkajorn, and J. Nukeawa, "Cell phone-based two-dimensional spectral analysis for banana ripeness estimation" Sensors and Actuators B **168**, 390 (2012).
14. S. Sumriddetchkajorn, K. Chaitavon, and Y. Intaravanne, "Mobile-platform based colorimeter for monitoring chlorine concentration in water" Sensors and Actuators B **191**, 561 (2014).
15. A. W. Martinez, S. T. Phillips, E. Carrilho, S. W. Thomas, H. Sindi, and G. M. Whitesides, "Simple telemedicine for developing regions: camera phones and paper-based microfluidic devices for real-time, off-site diagnosis" Anal. Chem. **80**, 3699 (2008).
16. A. García, M. M. Erenas, E. D. Marinetto, C. A. Abada, I. O. Paya, A. J. Palma, and L. F. C. Vallvey, "Mobile phone platform as portable chemical analyzer" Sensors and Actuators B **156,** 350 (2011).
17. J. L. Delaney, E. H. Doeven, A. J. Harsant, and C. F. Hogan, "Use of a mobile phone for potentiostatic control with low cost paper-based microfluidic sensors" Anal. Chim. Acta **790,** 56 (2013).
18. J. L. Delaney, C. F. Hogan, J. Tian, and W. Shen, "Electro-generated chemiluminescence detection in paper-based microfluidic sensors" Anal. Chem. **83**, 1300 (2011).
19. J. L. Delaney, E. H. Doeven, A. J. Harsant, and C. F. Hogan, "Use of a mobile phone for potentiostatic control with low cost paper-based microfluidic sensors" Anal. Chem. Acta **790**, 56 (2013).
20. J. Canning, A. Lau, M. Naqshbandi, I. Petermann, and M. J. Crossley, "Measurement of fluorescence in a rhodamine-123 doped self-assembled "giant" mesostructured silica sphere using a smartphone as optical hardware" Sensors **11**, 7055 (2011).
21. Mobile-cellular subscriptions 2013, [online]: http://www.itu.int/en/ITU-D/Statistics /Pages/stat/default.aspx.
22. B. Oram, Water Research Centre, [online]: http://www.water-research.net/index.php/ph-in-the-environment.
23. *GPS Visualizer*, [online]: http://www.gpsvisualizer.com/.
24. A. P. de Silva, H. Q. N. Gunaratne, J. L. Habib-Jiwan, C. P. McCoy, T. E. Rice, and J.-P. Soumillion, "New fluorescent model compounds for the study of photoinduced electron transfer: The influence of a molecular electric field in the excited state" Angew. Chem. Int. Ed. Eng. **34**, 1728 (1995).